# Relativistic dynamics without conservation laws

Bernhard Rothenstein[1), Stefan Popescu [2)

1) Politehnica University of Timisoara, Physics Department, Timisoara, Romania
2) Siemens AG, Erlangen, Germany

*Abstract. We show that relativistic dynamics can be approached without using conservation laws (conservation of momentum, of energy and of the centre of mass). Our approach avoids collisions that are not easy to teach without mnemonic aids. The derivations are based on the principle of relativity and on its direct consequence, the addition law of relativistic velocities.*

## 1. Introduction

Classical dynamics teaches us that a particle with mass $m_0$ moving with velocity **u** has a momentum **p**

$$\mathbf{p} = m_0 \mathbf{u} \tag{1}$$

and a kinetic energy

$$T = \frac{1}{2} m_0 u^2 . \tag{2}$$

Einstein's dynamics teaches us that the momentum of a particle is

$$\mathbf{p} = \frac{m_0 \mathbf{u}}{\sqrt{1 - \frac{u^2}{c^2}}} \tag{3}$$

whereas its kinetic energy is

$$T = m_0 c^2 \left( \frac{1}{\sqrt{1 - \frac{u^2}{c^2}}} - 1 \right) \tag{4}$$

where as we see $m_0$ represents the same Newtonian mass, relativists calling it rest mass, which accounts for the inertia of the particle at the moment when its acceleration starts from a state of rest. Relativists also introduce the concept of relativistic mass

$$m = \frac{m_0}{\sqrt{1 - \frac{u^2}{c^2}}} \tag{5}$$

of rest energy

$$E_0 = m_0 c^2 \tag{6}$$

and of relativistic energy

$$E = \frac{m_0 c^2}{\sqrt{1 - \frac{u^2}{c^2}}} = T + E_0 . \tag{7}$$

The art of the one teaching the relativistic dynamics consists in offering a justification for (3),(4),(5) and (6) which have a very good experimental confirmation. The job is done in different ways:
(1) Requiring momentum conservation in all inertial reference frames for collisions[1,2,3,4,5,6,7,8,9,10,11]. However, teaching relativistic dynamics by this way without mnemonic aids is not an easy task.
(2) Involving plane electromagnetic waves or photons in the derivations. Classical electrodynamics teaches us that that a plane electromagnetic wave carries energy $E$ and momentum $\mathbf{p}$ related by
$$\mathbf{p} = \frac{E}{c^2}\mathbf{c}. \qquad (8)$$
In Einstein's thought experiment, a body at rest in a given inertial reference frame simultaneously emits a plane electromagnetic wave of energy $\frac{E}{2}$ at an angle $\phi$ with the OX axis and another plane electromagnetic wave of equal energy in the opposite direction[12]. A similar approach is followed by other authors[13,14].
(3) As a result of guesswork or genial intuition, Einstein replaced the classical expression of the momentum
$$p = m_0 \frac{dx}{dt} \qquad (9)$$
with
$$p = m_0 \frac{dx}{d\tau} \qquad (10)$$
where $d\tau$ represents the proper (or "wristwatch" time). Because in accordance with the time dilation effect
$$dt = \frac{d\tau}{\sqrt{1-\frac{u^2}{c^2}}} \qquad (11)$$
the relativistic momentum becomes
$$\mathbf{p} = \frac{m_0 \mathbf{u}}{\sqrt{1-\frac{u^2}{c^2}}}. \qquad (12)$$
If an alert student would ask why we do that, it is not easy to find out a satisfactory answer[15,16].
(4) Many papers are devoted to the derivation of the mass-energy equivalence. The derivations involve the concept of force and the work energy theorem[17,18,19].
(5) Using the four vectors approach. It operates with the concept of proper time and proper mass. As we will show later on, the approach we propose leads in a natural way to the components of a four vector[20,21].

The derivation of (5) does not involve conservation laws. Tsai[22] derives it considering a thought experiment using a balance whose arm length can be varied in order to prove that both inertial masses of a moving body obey the relationship (5) within the scope of the special theory of relativity. The derivation involves the first principle (a balance in equilibrium in a given inertial reference frame is in the same state in all other inertial reference frames) and the composition law of relativistic velocities. Equation (5) is also the consequence of a Bucherer type experiment, putting in accordance the experimental results with the expected results on the basis of classical dynamics[23].

The inertial reference frames we involve are **K**(XOY) and **K'**(X'O'Y'). The corresponding axes of the two reference frames are parallel to each other and the OX(O'X') are common. At the origin of time in the two frames ($t = t' = 0$) the origins O and O' are located at the same point in space. Reference frame K' moves with constant velocity $V$ in the positive direction of the common axes. Consider a particle that moves with velocity $\mathbf{u}(u_x, u_y)$ relative to **K** and with velocity $\mathbf{u'}(u'_x, u'_y)$ relative to K'. At $t = t' = 0$ the particle goes through the point where the origins O and O' of the two frames are located. When detected from **K** the particle moves along a direction that makes an angle $\theta$ with the positive direction of the OX axis given by

$$\tan\theta = \frac{u_y}{u_x}. \tag{13}$$

The same angle detected from **K'** is $\theta'$ given by

$$\tan\theta' = \frac{u'_y}{u'_x}. \tag{14}$$

Relativistic kinematics teaches us that the components of $\mathbf{u}$ and $\mathbf{u'}$ are related by

$$u_x = \frac{u'_x + V}{1 + \frac{Vu'_x}{c^2}} = u'\frac{\cos\theta' + \frac{V}{u'}}{1 + \frac{Vu'\cos\theta'}{c^2}} \tag{15}$$

$$u_y = \frac{u'_y\sqrt{1 - \frac{V^2}{c^2}}}{1 + \frac{Vu'_x}{c^2}} = u'\frac{\sqrt{1 - \frac{V^2}{c^2}}\sin\theta'}{1 + \frac{Vu'\cos\theta'}{c^2}} \tag{16}$$

the magnitudes of the velocities being related by

$$u = u'\frac{\sqrt{(\cos\theta' + V/u')^2 + (1 - V^2/c^2)\sin^2\theta'}}{1 + \frac{Vu'\cos\theta'}{c^2}}. \tag{17}$$

Equations (15) and (16) represent the addition law of relativistic velocities and are usually derived using the Lorentz-Einstein transformations for the space-time coordinates of the same event[24]. We can derive them even without using the Lorentz-Einstein transformations.[25,26,27,28] The angles $\theta$ and $\theta'$ are related by

$$tan\,\theta = \frac{\sqrt{1-V^2/c^2}\,sin\,\theta'}{cos\,\theta' + \frac{V}{u'}} \tag{18}$$

The inverse transformation can be obtained by changing in the equations derived above the corresponding primed physical quantities with un-primed ones and changing the sign of *V*. Equations (15) and (16) enable us to construct the following relativistic identities

$$(1-u^2/c^2)^{-1/2} = (1-u'^2/c^2)^{-1/2}(1-V^2/c^2)^{-1/2}(1+\frac{Vu'_x}{c^2}) \tag{19}$$

$$u_x(1-u^2/c^2)^{-1/2} = (1-u'^2/c^2)^{-1/2}(1-V^2/c^2)^{-1/2}(u'_x+V) \tag{20}$$

$$u_y(1-u^2/c^2)^{-1/2} = u'_y(1-u'^2/c^2)^{-1/2}. \tag{21}$$

In our derivations we will exploit the fact that (19),(20) and (21) remain identities if we multiply both their sides by an invariant physical quantity (proper time, proper mass).

**The purpose of our paper is to show that the relativistic dynamics could be approached without using conservation of momentum, of relativistic mass (energy) or of the centre of mass but by using instead the composition law of relativistic velocities.**

**2. Relativistic dynamics without conservation laws**
**2.1 Mimicking classical dynamics**

When detected from **K** the particle we have mentioned above has the momentum $\mathbf{p}(p_x, p_y)$ whose components are

$$p_x = mu_x \tag{22}$$
$$p_y = mu_y. \tag{23}$$

Detected from **K'** the momentum is $\mathbf{p}'(p'_x, p'_y)$ having the components

$$p'_x = m'u'_x \tag{24}$$
$$p'_y = m'u'_y. \tag{25}$$

in accordance with the first postulate.

In classical dynamics we consider $m = m'$. Combining (22) and (24) we obtain

$$p_x = p'_x \frac{u_x}{u'_x} = p'_x + mV \tag{26}$$

where we have used the classical composition law of velocities. We extend the same procedure to the case of relativistic dynamics. Combining (22) and (24) we obtain

$$\frac{p_x}{m} = \frac{p'_x}{m'}\frac{u_x}{u'_x} = \frac{p'_x}{m'}\frac{1+V/u'_x}{1+Vu'_x/c^2} \qquad (27)$$

where we have taken into account the addition law of relativistic velocities (15). Linearity suggests considering that we should have

$$p_x = F'(V)p'_x(1+V/u'_x) = F'(V)(p'_x + Vm') \qquad (28)$$

and

$$m = F'(V)m'(1+Vu'_x/c^2) = F'(V)(m' + Vp'_x/c^2) \qquad (29)$$

In a similar way we obtain

$$\frac{p'_x}{m'} = \frac{p_x}{m}\frac{1-V/u_x}{1-Vu_x/c^2} \qquad (30)$$

suggesting that we should have

$$p_x = F(V)p_x(1-V/u_x) \qquad (31)$$
$$m' = F(V)m(1-Vu_x/c^2) \qquad (32)$$

*F(V)* and *F'(V)* representing unknown functions of the relative velocity *V* but not of the physical quantities involved in the transformation process.

Consider (29) in the particular case when $u'_x = 0$. Under such conditions observers from **K'** measure its rest mass $m_0$ observers from **K** measuring its mass $m_{u'_x=0}$ given by

$$m_{u'_x=0} = F'(V)m_0. \qquad (33)$$

Consider (32) in the particular case when $u_x = 0$. Under such conditions the particle is at rest in **K**, observers of that frame measure its rest mass, observers from **K'** measuring its mass $m'_{u_x=0}$ given by

$$m'_{u_x=0} = F(V)m_0. \qquad (34)$$

The first principle requires that $m_{u'_x=0} = m'_{u_x=0}$ and so we have

$$F(V) = F'(V). \qquad (35)$$

Multiplying (29) and (32) side by side we obtain taking into account (35)

$$1 = F^2(V)(1+Vu'_x/c^2)(1-Vu_x/c^2). \qquad (36)$$

Expressing the right hand side of (36) as a function of $u_x$ or as a function of $u'_x$ only via the addition law of relativistic velocities simple algebra leads to

$$F(V) = F'(V) = (1-V^2/c^2)^{-1/2} = \gamma \qquad (37)$$

and so we obtain the transformation equations

$$p_x = \gamma p'_x(1+V/u'_x) = \gamma(p_x + Vm') \qquad (38)$$
$$m = \gamma m'(1+Vu'_x/c^2) = \gamma(m' + Vp'_x/c^2). \qquad (39)$$

We obtain the transformation equation for the OY(O'Y') component of the momentum starting with (23) and expressing its right hand side as a function of physical quantities measured from **K'**

$$p_y = mu_y = m'u'_y = p'_y. \qquad (40)$$

As we see, the OY(O'Y') component of the momentum is a relativistic invariant.

The magnitude of the momentum transforms as

$$p = \sqrt{\gamma^2(p'_x + Vm')^2 + p'^{2}_y}. \qquad (41)$$

If the particle is at rest in **K'** ($p'_x = 0, p'_y = 0$) observers from **K'** measure its rest mass $m_0$, (40) leading to

$$\mathbf{p} = \gamma m_0 \mathbf{V} \qquad (42)$$

relativists calling **p** relativistic momentum of a particle with rest mass $m_0$ moving with velocity **V**. Under the same condition (40) leads to the expression for the relativistic mass

$$m = \gamma m_0 \qquad (43)$$

of a particle with rest mass $m_0$ moving with speed V.

**2.2 Involving relativistic identities and the concept of invariant rest mass $m_0$**

We start with the relativistic identities (19),(20) and (21), derived above, multiplying both theirs sides with $m_0$. The result is

$$\left[\frac{m_0}{\sqrt{1-u^2/c^2}}\right] = \frac{\left[\frac{m_0}{\sqrt{1-u'^2/c^2}}\right] + \frac{V^2}{c^2}\left[\frac{m_0 u'_x}{\sqrt{1-u'^2/c^2}}\right]}{\sqrt{1-V^2/c^2}} \qquad (44)$$

$$\left[\frac{m_0 u_x}{\sqrt{1-u^2/c^2}}\right] = \frac{\left[\frac{m_0 u'_x}{\sqrt{1-u'^2/c^2}}\right] + V\left[\frac{m_0}{\sqrt{1-u'^2/c^2}}\right]}{\sqrt{1-V^2/c^2}} \qquad (45)$$

$$\left[\frac{m_0 u_y}{\sqrt{1-u^2/c^2}}\right] = \left[\frac{m_0 u'_y}{\sqrt{1-u'^2/c^2}}\right]. \qquad (46)$$

Equations (44),(45) and (46) are genuine transformation equations per se. Physicists, as well trained godfathers, find out names for the physical quantities in the brackets, in concordance with theirs physical dimensions, and recover the transformation equations derived above.

**2.3 Involving (5) derived without conservation laws**

The relativistic mass of the particle, detected by observers from **K** is

$$m = \frac{m_0}{\sqrt{1-u^2/c^2}}. \qquad (47)$$

In accordance with the first postulate, the relativistic mass of the same particle, as detected from **K'** is

$$m' = \frac{m_0}{\sqrt{1-u'^2/c^2}}. \tag{48}$$

Eliminating $m_0$ between (47) and (48) and taking into account (17) we obtain that *m* and *m'* are related by

$$m = \frac{m' + Vm'u'_x/c^2}{\sqrt{1-V^2/c^2}}. \tag{49}$$

We detect in the right side of (48) the presence of the term

$$p'_x = m'u'_x = \frac{m_0 u'_x}{\sqrt{1-u'^2/c^2}}. \tag{50}$$

It has the physical dimensions of relativistic momentum and relativists call it the O'X' component of the momentum. Detected from K it is

$$p_x = mu_x = \frac{m_0 u_x}{\sqrt{1-u^2/c^2}}. \tag{51}$$

Eliminating $m_0$ between (49) and (50) we recover (38).

**3. Introducing the concept of relativistic energy and putting the ban on the concept of relativistic mass**

We start with the obvious identity

$$(1-u^2/c^2)^{-1/2} - (1-u^2/c^2)^{-1/2} u^2/c^2 = 1. \tag{52}$$

Multiplying both its sides with $m_0^2 c^4$ it becomes the relativistic identity

$$m_0^2 c^4 (1-u^2/c^2)^{-1/2} - m_0^2 c^2 (1-u'^2/c^2)^{-1/2} = m_0^2 c^4. \tag{53}$$

We recognise that the term

$$m_0^2 u^2 (1-u^2 c^{-2})^{-1/2} = p^2 \tag{54}$$

represents the square of the momentum in **K.** But what is the physical meaning of the first term. We see that it has the physical dimensions of energy and presenting it as

$$E = m_0 c^2 (1-u^2 c^{-2})^{-1/2} = m_0 c^2 + \frac{m_0 u^2}{2} + \cdots = mc^2 \tag{55}$$

it contains in its power development the classical kinetic energy. Relativists call it relativistic energy E, a generally accepted concept. With the new notations (53) becomes

$$E^2 c^{-2} - p_x^2 - p_y^2 = m_0^2 c^2. \tag{56}$$

Equation (56) represents an invariant combination between the momentum and the energy of the same particle.

Expressed as a function of the energy the OX(O'X') component of the momentum transforms as (38)

$$p_x = \frac{p'_x + Vc^{-2} E'}{\sqrt{1-V^2/c^2}}. \tag{57}$$

Multiplying both sides of (49) with $c^2$ we obtain the transformation equation for energy

$$E = \frac{E' + Vp'_x}{\sqrt{1 - V^2/c^2}}. \tag{58}$$

If the particle is at rest in **K'**, observers of that frame measure its rest energy

$$E_0 = m_0 c^2 \tag{59}$$

Whereas observers from **K** measure its energy

$$E = \frac{m_0 c^2}{\sqrt{1 - V^2/c^2}} \tag{60}$$

because the particle moves relative to them with speed *V*.

The single supplementary energy a free electron can possess is its kinetic energy *T* that should equate the difference between its energy *E* and its rest energy i.e.

$$T = E - E_0 = E_0 \left( \frac{1}{\sqrt{1 - u^2/c^2}} - 1 \right) \tag{61}$$

In the reference frame relative to which the electron moves with speed *V*.

The result is that we can put the ban on the concept of relativistic mass[29] characterizing a particle by its rest mass $m_0$, by its rest energy $E_0$, by its energy *E* and by its momentum

$$\mathbf{p} = \frac{E}{c^2} \mathbf{u} = \frac{m_0 \mathbf{u}}{\sqrt{1 - u^2/c^2}} \tag{62}$$

of course in the reference frame relative to which it moves with speed u. There are authors who defend the concept of relativistic mass[30], others having a conciliatory attitude[31].

Presenting (55) as

$$\Delta E = c^2 \Delta m \tag{63}$$

it accounts for Einstein's mass-energy equivalence, telling us that a change of the mass of a particle is accompanied by a change in its energy and vice versa[32].

**4. The four vector approach**

The particle we have considered so far generates after a given time of motion the event $M(x = r\cos\theta, y = r\sin\theta, t = r/u)$, x and y representing the Cartesian space coordinates, $r = \sqrt{x^2 + y^2}$, and $\theta$ the polar coordinates and $t = r/u$ the time coordinate. A clock commoving with the particle measures the invariant proper time interval $\tau$ related to *t* by the time dilation formula

$$t = \frac{\tau}{\sqrt{1 - \frac{u^2}{c^2}}}, \tag{64}$$

taking into account that at the origin of time the particle was located at the origin O. Starting with the identity (52) and multiplying both its sides with $c^2\tau^2$ it leads to

$$\frac{c^2\tau^2}{1-u^2/c^2} - \frac{u^2\tau^2}{1-u^2/c^2} = c^2\tau^2 \qquad (65)$$

or in a three space dimensions approach

$$c^2t^2 - x^2 - y^2 - z^2 = c^2\tau^2 \qquad (66)$$

which establishes a reference frame independent relationship between the space-time coordinates of event *M*. Relativistic kinematics teaches us that the identity is satisfied if the space-time coordinates of the same event in **K** and in **K'** are related by the Lorentz-Einstein transformations

$$x = \frac{x' + Vt'}{\sqrt{1-V^2/c^2}} \qquad (67)$$

$$y = y' \qquad (68)$$

$$z = z' \qquad (69)$$

$$ct = \frac{ct' + Vc^{-1}x'}{\sqrt{1-V^2/c^2}}. \qquad (70)$$

Multiplying both sides of (52) with the rest mass of the particle $m_0$ it becomes

$$\left[\frac{m_0^2}{1-u^2/c^2}\right] - \frac{1}{c^2}\left[\frac{m_0^2 u^2}{1-u^2/c^2}\right] = m_0^2. \qquad (71)$$

Finding out names for the physical quantities in the brackets in accordance with theirs physical dimensions we can present (71) as

$$m^2 - c^{-2}(p_x^2 + p_y^2 + p_z^2) = m_0^2 c^2 \qquad (72)$$

or as

$$m^2 c^4 - c^2(p_x^2 - p_y^2 - p_z^2) = m_0^2 c^4. \qquad (73)$$

Equations (67)-(70) enable us to define a position four vector **R**(x,y,z,ct) in **K** and **R'**(x',y',z',ct') in **K'**, taking into account the definition of a four vector as any quantity of four components which transform under a Lorentz transformation like the four components of the space time coordinates of an event do. Because $E = mc^2$ has the physical dimensions of energy we can present (73) as

$$E^2 - c^2(p_x^2 + p_y^2 + p_z^2) = E_0^2 \qquad (74)$$

The algebraic structure of (72) suggests considering that *m* should transform as the time coordinate does i.e.

$$m = \frac{m' + Vc^{-2}p'_x}{\sqrt{1-V^2/c^2}} \qquad (75)$$

or

$$E = \frac{E' + Vp'_x}{\sqrt{1-V^2/c^2}} \qquad (76)$$

The components of the momentum transform as

$$p_x = \frac{p'_x + Vm'}{\sqrt{1-V^2/c^2}} = \frac{p'_x + Vc^{-2}E'}{\sqrt{1-V^2/c^2}} \tag{77}$$

$$p_y = p'_y \tag{78}$$

and

$$p_z = p'_z. \tag{79}$$

We can define the momentum four vector as $\mathbf{P}(p_x, p_y, p_z, E/c)$ in the **K** reference frame.

Our approach to the concept of momentum four vector is straightforward and no less transparent then the usual approach which performs successive derivations of the space-time coordinates of an event with respect to the proper time interval.

**5. Involving the photon**

As we have mentioned above, many relativistic dynamics derivations involve the photon, a particle that exists only moving with the invariant velocity **c**. Electrodynamics teaches us, without involving special relativity, that an electromagnetic wave, propagating in empty space, carries energy $E$ and momentum $\mathbf{p} = \frac{E}{c^2}\mathbf{c}$. Let $N$ be the invariant number of photons which carry the energy $E$. The result is that a single photon carries energy $E_{ph}$ and a momentum $p_{ph}$ related by

$$p_{ph} = \frac{E_{ph}}{c}. \tag{80}$$

Einstein expresses the energy of a photon as

$$E_{ph} = h\nu \tag{81}$$

and its momentum as

$$p_{ph} = \frac{h\nu}{c} \tag{82}$$

$\nu$ representing the frequency of the electromagnetic oscillations taking place in the electromagnetic wave to which we attach the photon. The invariance of $c$ tells us that in special relativity, energy and momentum transform via the same transformation factor $D_c$ i.e.

$$p_{ph} = D_c p'_{ph} \tag{83}$$

and

$$E_{ph} = D_c E'_{ph}. \tag{84}$$

Consider that a source **S'** at rest in **K'** and located at its origin **O'** emits a bullet with velocity $u'_x$ at $t'=0$ and a light signal under the same conditions, both in the positive direction of the common OX(O'X') axes. We present in Table 1 the transformation equations that relate the physical quantities introduced in order to describe the motion of the bullet (tardyon) and of the photon in **K** and in **K'** respectively.

| Bullet | Photon |
|---|---|
| $x = x' \dfrac{1+V/u'}{\sqrt{1-V^2/c^2}}$ | $x_c = x'_c \dfrac{1+V/c}{\sqrt{1-V^2/c^2}}$ |
| $t = t' \dfrac{1+Vu'/c^2}{\sqrt{1-V^2/c^2}}$ | $t_c = t'_c \dfrac{1+V/c}{\sqrt{1-V^2/c^2}}$ |

**Table 1.**

As we see, the transformation factors that perform the transformation in the case of the photon are the limits for u'=c of the factors that perform the transformation in the case of the tardyon.

Considering an acoustic wave detected from **K** and from **K'** respectively, propagating in the positive direction of the common axes. Moller[33] establishes that the frequencies of the mechanical oscillations taking place in the wave in the two frames, $\nu$ and $\nu'$ respectively are related by

$$\nu = \nu' \frac{1+Vu'/c^2}{\sqrt{1-V^2/c^2}} \tag{85}$$

u' representing the propagation velocity of the wave, as detected from **K'**. In the case of an electromagnetic wave (85) becomes

$$\nu_c = \nu'_c \frac{1+V/c}{\sqrt{1-V^2/c^2}} = \sqrt{\frac{1+V/c}{1-V/c}} \tag{86}$$

equations (83) and (84) becoming

$$p_{ph} = p'_{ph} \sqrt{\frac{1+V/c}{1-V/c}} \tag{87}$$

$$E_{ph} = E'_{ph} \sqrt{\frac{1+V/c}{1-V/c}}. \tag{88}$$

We can also consider that $u'$ in (85) represents the velocity of bullets successively emitted at a constant frequency $\nu$ in **K** and $\nu'$ in **K'**. If we postulate that **special relativity ensures a smooth transition from the equations that describe the behaviour of a tardyon to the behaviour of a photon** we could consider, taking into account (85), that the energy of a tardyon should transform as

$$E = E' \frac{1+Vu'/c^2}{\sqrt{1-V^2/c^2}} = \frac{E'+V(E'u'/c^2)}{\sqrt{1-V^2/c^2}}. \tag{89}$$

We detect in the right side of (89) the presence of the term

$$p'_x = E'u'/c^2 \tag{90}$$

that reads in **K'**

$$p_x = Eu/c^2 \tag{91}$$

having the expected limits for $u = u' = c$ and the physical dimensions of momentum.

Combining (90) and (91) we obtain

$$p_x = \frac{p'_x + VE'/c^2}{\sqrt{1-V^2/c^2}} \tag{92}$$

all the results being obtained without using conservation laws.

**Conclusions**

We have presented an approach to relativistic dynamics that avoids conservation laws (conservation of momentum, conservation of relativistic mass (energy) and conservation of the centre of mass) and collisions, which are not easy to teach at the blackboard without using mnemonic aids. Our approach also avoids the concept of force and the kinetic energy theorem. It implies a single scenario (a particle that moves along a given direction with the common OX(O'X') axes and avoids ad-hoc assumptions at different points of the derivation.

Evaluating the outcome of a lecture we should take into account the time invested and the understanding achieved. We consider that our approach satisfies both conditions.

**References**
[1]R.P. Feynman, R.B.Leighton, and M.Sands, *The Feynman lectures of Physics, Vol.I* (Reading, Ma. Addison-Wesley 1963) pp. 16/6-16/7)
[2]E.F. Taylor and J.A. Wheeler, *Space-time Physics* ( Freeman, New York 1966) pp.106-108
[3]Charles Kittel, Walter D. Knight and Malvin Ruderman, *Berkeley Physics Course Volume 1, Second Edition* (Mc Graw Hill, 1973) Ch.12.1
[4]Jay Orear, *Physik,* (Carl Hanser Verlag Muenchen, Wien, 1982) pp.172-173
[5]R. Breneke and A.Schuster, *Physik,* (Friedr Vieweg Sohn, Brunschweig 1966/1969) Ch. 8.5
[6]R.C. Tolman, *Relativity, Thermodynamics and Cosmology,* (reissued by Dover 1987) pp.43-44; G.H. Lewis and R.C. Tolman, "The principle of relativity and non-Newtonian mechanics," Philos. Mag. **18,** 510-523 (1909)
[7]D. Bohm, *The Special Theory of Relativity,* (Routledge, London) pp.84-88
[8]A. Shadowitz, *Special Relativity,* (N.B. Saunders, 1968) Ch.6
[9]Robert Resnick, *Introduction to Special Relativity,* (John Wiley and Sons, Inc. New York (1968) pp.110-119
[10]Sebastiano Sonego and Massimo Pin, "Deriving relativistic momentum and energy" Eur.J.Phys. **26,** 33-45 (2005)
[11]P.C. Peters, "An alternate derivation of the relativistic momentum" Am.J.Phys. **54,** 804-808 (1986) and references cited therein
[12]A. Einstein, "Does the inertia of a body depend on its energy content?, Ann. Phys. **18**, 639-641 (1905)
[13]L. Sartori, "On the derivation of the formula for relativistic momentum," Am.J.Phys. **62,** 280-281 (1994)
[14]Daniel J. Steck, "An elementary development of mass-energy equivalence," Am.J.Phys. **51,** 461-462 (1983)

[15] Jon Ogborn, "Introducing relativity: Less my be more," Phys.Educ. **40,** 213-222 (2005)

[16] N. David Mermin *It's About Time: Understanding Einstein's Relativity,"* (Princeton University Press Princeton and Oxford 2005) Ch.11

[17] Flores Francisco, "The equivalence of mass and energy," The Stanford Encyclopaedia of Philosophy (Fall 2004 Edition, Edward N. Zalta(ed)

[18] Adel F. Antippa, "Inertia of energy and the liberated photon," Am.J.Phys. **44**, 841-844 (1976)

[19] Mitchell J. Feigenbaum and N. David Mermin, "$E=mc^2$" Am.J.Phys. **56**, 18-22 (1988)

[20] N.D. Mermin, *Space and Time in Special Relativity* (Mc Graw Hill, reissued by Valeland Press 1989) pp.207-214

[21] Robert W. Resnick, "The advantage of teaching special relativity with four vectors," Am.J.Phys. **36**, 896-901 (1968)

[22] Ling Tsai, "The relation between gravitational mass, inertial mass, and velocity," Am.J.Phys.**54**, 340-342 (1986); Michel Nicola, Mass velocity dependence without the momentum principle," **45,** 1218 (1977)

[23] W.G.V. Rosser, *An Introduction to the Theory of Relativity,"* (Butherworth,London 1964) pp.193-195

[24] Thomas A. Moore, *A Traveler's Guide to Spacetime ,* (McGraw-Hill,Inc. New York, 1995) pp.148-150

[25] L.Sartori, "Elementary derivation of the relativistic addition law," Am.J.Phys. **63,** 81-82 (1995)

[26] Asher Peres, "Relativistic telemetry," Am.J.Phys. **55,** 516-519

[27] Margaret Stautberg Greenwood, "Relativistic addition of velocities using Lorentz contraction and time dilation," Am.J.Phys. **50,** 1156-1157 (1982)

[28] N. David Mermin, "Relativistic addition of velocities directly from the constancy of the velocity of light," Am.J.Phys. **51**, 1130-1131 (1983)

[29] L.B. Okun, "The concept of mass," Phys.Today **42,** 31-36 (1989)

[30] T.R. Sadin, "In defence of relativistic mass," Am.J.Phys. **59,** 1032 (1991)

[31] R.P. Bikerstaff and G. Patsakos, "Relativistic generalisations of mass," Eur.J.Phys. **16**, 63-68 (1996)

[32] E.G. Thomas, "What's so special about $E=mc^2$," European Journal of Physics **26**, S125-S130 (2005)

[33] C. Moller, *The Theory of Relativity,* (Clarendon Press, Oxford, 1972) Ch.2.9